\documentclass[11pt]{article}

\usepackage[usenames,dvipsnames]{color}
\usepackage{amsmath}
\usepackage{amssymb}
\usepackage[numbers,square,sort&compress]{natbib}
\usepackage{listings}
\usepackage{graphicx}
\usepackage{fullpage}

\usepackage[colorlinks, bookmarksopen, 
            linkcolor={black},
            anchorcolor={black},
            citecolor={black},
            filecolor={magenta},
            menucolor={blue},
            pagecolor={red},
            plainpages=false,pdfpagelabels,
            urlcolor={black}]{hyperref}

\usepackage{subfig}
\usepackage{tikz}
\usetikzlibrary{positioning,arrows,decorations,backgrounds,shapes}
\tikzstyle{expr}=[circle,draw=black]
\tikzstyle{dofexpr}=[rectangle,draw=black]
\tikzstyle{dep}=[thick,draw=black,-stealth']
\tikzstyle{dep2}=[thick,dashed,draw=red,-stealth']
\tikzstyle{dep3}=[thick,dotted,draw=blue,-stealth']

\providecommand\bcdot{\boldsymbol{\cdot}}

\providecommand\bnabla{\boldsymbol{\nabla}}

\renewcommand{\vector}[1]{\ensuremath{{\boldsymbol{#1}}}}

\newcommand{\diff}{\ensuremath{\mathrm{d}}}

\newcommand{\grad}{\ensuremath{\bnabla}}

\newcommand{\Vol}{\ensuremath{\Omega}}

\newcommand{\dV}{\ensuremath{\, \diff \Vol}}

\DeclareGraphicsExtensions{.pdf,.png,.jpg}

\begin{document}

\title{Automating embedded analysis capabilities and managing software complexity in multiphysics simulation part II:  application to partial differential equations}
\author{Roger P. Pawlowski, Eric T. Phipps, Andrew G. Salinger\footnote{Corresponding author. Sandia National Laboratories, Numerical Analysis and Applications Department, PO Box 5800 MS-1318, Albuquerque, New Mexico, 87185, USA. Tel: +1(505)845-3523, Fax: +1(505)845-7442, {\tt agsalin@sandia.gov}.},\\ Steven J. Owen, Christopher M. Siefert, and Matthew L. Staten, \\
Sandia National Laboratories\footnote{Sandia National Laboratories is a multi-program laboratory managed and operated by Sandia Corporation, a wholly owned subsidiary of Lockheed Martin Corporation, for the U.S. Department of Energy's National Nuclear Security Administration under contract DE-AC04-94AL85000.}}
\maketitle

\begin{abstract}
A template-based generic programming approach was presented in a previous paper~\cite{PPS:2011}
that separates the development effort of
programming a physical model from that of computing additional quantities, such as derivatives,
needed for embedded analysis algorithms. In this paper, we describe the implementation
details for using the template-based generic programming approach for simulation and analysis of partial differential
equations (PDEs). We detail several of the hurdles that we have encountered, and some of
the software infrastructure developed to overcome them. We end with a demonstration
where we present shape optimization and uncertainty quantification results for
a 3D PDE application.
\end{abstract}

\section{Introduction}
Computational science has the potential to provide much more than numerical
solutions to a set of equations.
The set of analysis opportunities beyond simulation include
parameter studies, stability analysis, optimization, and uncertainty quantification.
These capabilities demand more from the application code than required for a single simulation, typically in the
form of extra derivative information. In addition, computational design and analysis
will often entail modification of the governing equations, such as
refinement of a model or a hierarchy of fidelities.

In our previous paper~\cite{PPS:2011}, we described the template-based generic
programming (TBGP) approach. 
That paper provides the conceptual framework upon which this paper builds, and thus is a prerequisite for the work described here.
We showed how  graph-based assembly and
template-based automatic differentiation technology can work
together to deliver a flexible assembly engine, where model equations can be
rapidly composed from basic building blocks and where only the residual
needs to be explicitly programmed.  The approach is based on templating of the low-level scalar operations within a simulation and instantiation of this template code on various data types to effect the code transformations needed for embedded analysis through operator overloading.  Often application of operator overloading in this manner is assumed to introduce significant run-time overhead into the simulation, however we have demonstrated~\cite{PPS:2011} that careful implementation of the overloaded operators~\cite{Phipps2012LST} (using techniques such as expression templates~\cite{VelhuizenET}) can completely eliminate this overhead.  This results in a single templated code base that must be developed, tested, and maintained,\footnote{We note that transforming a legacy implementation to use templates in this manner does involve significant effort (and thus we would consider this approach most appropriate for new development efforts), however the transformations necessary are just type specifications in function and variable declarations.} that when combined with 
appropriate seeding and extracting
of these specially-designed overloaded data types (see~\cite{PPS:2011} and Section~\ref{sect:pde} for definitions of these terms), 
allows all manner of additional quantities
to be generated with no additional software development time.   

In this paper, we extend the description of this approach to the
simulation and analysis of partial differential equations (PDEs). As
discussed in the previous paper \cite{PPS:2011}, a number of projects
have implemented embedded analysis capabilities that leverage a domain
specific language.  Specifically for finite elements, the FEniCS
\cite{FenicsOverview2007,LoggMardalEtAl2012a}, Life/FEEL++
\cite{Life2006,Life2006b} and Sundance \cite{Long2010} projects have
demonstrated this capability with respect to derivative evlauation.

 PDEs
provide additional challenges with regards to data structures and
scalability to large systems. In this paper, we deal specifically
with a Galerkin finite element approach, though the approach will
follow directly to other element-based
assemblies, and by analogy to stencil-based assemblies.
In Section \ref{sect:fem} we discuss where the template-based
approach begins and ends, and how it relates to the global
and local (element-based) data structures.
In Section~\ref{sect:pde} we present many details of the
template-based approach for finite element assembly, in
particular the seed, compute, and extract phases. 
Section~\ref{sect:pdefill} addresses some more advanced issues
that we have dealt with in our codes that use this approach.
Specifically, this includes the infrastructure for exposing
model parameters, as needed for
continuation, bifurcation, optimization and uncertainty quantification, approaches for dealing
with a templated code stack, and
approaches for dealing with code
that can not be templated.
Finally, in Section~\ref{sect:app} we demonstrate the
whole process on an example PDE application: the
sliding electromagnetic contact problem. We show results
for shape optimization and embedded uncertainty quantification.

Critical to the main message of this paper is the fact
that the infrastructure for computing the extra quantities
needed for these analysis
capabilities has been implemented independently from the
work of implementing the PDE model. This infrastructure includes the
seed and extract phases for the template-based approach. It also includes all of
the solver libraries that have been implemented in the Trilinos
framework \cite{trilinosTOMS}, such as the linear, nonlinear,
transient, optimization, and UQ solvers.
Once in place, application codes for new PDEs
can be readily generated, born with analytic
derivatives and embedded analysis capabilities.

The novel interface exposed to computational scientists
by allowing for templated data types to be passed through the
equation assembly has tremendous potential. While this interface
has been exploited for derivatives,
operation counting, and polynomial propagation, we expect
that developers will find innovative ways to exploit this
interface beyond what we currently imagine.

\section{Approach for Finite Element Codes}\label{sect:fem}

In the first paper in this series, we explained the
template-based generic programming approach and included an
illustrative demonstration
on how it can be applied to an ODE problem. 
In this section, we present the basic details on
how this approach is used in the context of PDE applications. Some of
our implementation details are restricted to discretization
strategies with element-based assembly kernels, such as finite element (FEM)
and control volume finite element (CVFEM) methods. Some details
of the approach would need to be adapted for stencil-based discretizations,
such as finite difference methods or integral equations, or for
discontinuous-Galerkin methods.

Extending the approach from ODEs to PDEs gives rise to many issues.
The core design principle is still the same, that the evaluation of
the equations is separated into three phases: seed, compute, and
extract. The seed and extract phases need to be specialized for each
template type,
where extra information in the data types (such as derivative information)
must be initialized and retrieved. The compute phase, where the equations
are implemented, can
be written on a fully generic fashion. There are also issues with regard to
data structures, sparse matrices, parallelism,
the use of discretization libraries, and the potential
dependency on libraries for property data. These issues will be
addressed in the following sections.

\subsection{Element-Based Assembly}
A primary issue that arises when using the template-based generic
programming (TBGP) approach
for PDEs is the sparsity of the derivative dependencies.
The automatic differentiation approach to computing the Jacobian
matrix using the Sacado package requires all relevant variables
to be a Sacado::FAD (forward automatic differentiation) data type, which includes
a dense array of partial derivatives with respect to
the independent variables in the problem. As problem sizes can
easily extend into
to the millions and beyond, yet nonzero entries per row stay bounded
at $O(100)$, it is not feasible to adopt the same approach.
A second issue is the requirement for the ability to
run the codes on distributed-memory parallel architectures.
Adding message-passing layers within the AD infrastructure would
also be challenging.

These two issues are circumvented by invoking the template-based
generic programming at a {\em local} level. For FEM methods, this
is the single element. The entire PDE assembly phase is performed
by summing contributions over
individual elements. Within each element,
it is typically not a bad assumption that the local Jacobian
(often referred to as the element stiffness matrix) is dense.
So, for Jacobian matrices, the AD is performed at the element
level, where the array of partial derivatives is sized to be the
number of degrees of freedom in an element. The dense contributions
to each row of the matrix is subsequently scattered to the global sparse
matrix structure.
Similarly, other quantities that can be computed with the
template-based generic programming
approach can also be calculated element by element,
and summed into a global data structure.\footnote{Note that most AD tools including Sacado compute the residual along with the Jacobian allowing these quantities to be computed simultaneously.  In general evaluation of the $n$th derivative also involves simultaneous evaluation of derivatives of order 0 up to $n-1$ as well.}

The choice of implementing the template-based generic programming
at a local level also nullifies the second issue
to do with distributed memory parallelism. In a typical
distributed memory implementation, information from neighboring
elements (often called ghost, overlap, or halo data) is
pre-fetched. The templating
infrastructure and seed-compute-extract loop
falls below the message-passing layer. In our implementation,
no communication is performed within the templated code.

We note that the local element-based approach is not the only solution to these problems.  Through the use of sparse derivative arrays or graph-based compression techniques (see~\cite{Griewank2000EDP} for an overview of both of these approaches and references to the relevant literature) automatic differentiation can be applied directly at the global level.  Furthermore, message passing libraries for distributed memory parallelism can be augmented to support communication of derivative quantities.  However, due to the extra level of indirection introduced, the use of sparse derivative arrays can significantly degrade performance.  Moreover compression techniques require first computing the derivative sparsity pattern and then solving an NP-hard optimization problem to compress the sparse derivative into a (nearly) dense one.  In practice only approximate solutions to this optimization problem can be attained.  However the solution to this problem is in fact known a priori, it is precisely equivalent to the local element-based approach (assuming the element derivative is dense).  Thus we have found the local element-based approach to be significantly simpler than a global one, particularly so as PDE discretization software tools that support templated data types have been developed, such as the Intrepid package in Trilinos.

\subsection{Data Structures}

The purpose of the PDE assembly engine is to fill
linear algebra objects -- primarily vectors and sparse matrices.
These are the data structures used by the solvers and analysis
algorithms. For instance, a Newton based solver will need a residual
vector and a Jacobian matrix; a matrix-free algorithm will
need a Jacobian-vector directional derivative; an explicit
time integration algorithm will need the forcing vector $f(x)$;
polynomial chaos propagation~\cite{Wiener:1938p989,Ghanem:1990p7167,Ghanem_Spanos_91,Xiu:2002p919} creates a vector of vectors of
polynomial coefficients; a sensitivity solve computes
multiple vectors (or a single multi-vector or dense column
matrix) of derivatives with respect to a handful of
design parameters. The input to the PDE assembly is also
vectors: the solution vector $x$, a vector of design
parameters $p$, and coefficient vectors for polynomial expansions
of parameters $\xi$.

\begin{table}
  \begin{center}
    \caption{Embedded analysis algorithms require a variety of quantities
    to be computed in the PDE assembly. This table shows a list of
    linear algebra quantities that can be computed, as well as the
    required inputs. In this table: $x$ is the solution vector, $\dot{x}$ is the time derivative of $x$,
    ${v}$ is one or more vectors chosen by the analysis algorithm, ${p}$ is one or more system parameters for continuation/optimization, ${\xi}$ is one or more random variables, $f$ is the residual vector of the discretized PDE system, and $F$ is the stochastic expansion of the residual vector.}
    \label{tab:embedded}
    \bigskip
    \begin{tabular}{|l|c|c||c|c|}
      \hline
      Evaluation & Input Vector(s) & Other Input & Output Vector & Output Matrix \\
      \hline
      Residual & $x$ & ${p}$ & $f$ &  \\
      Steady Jacobian & $x$ & ${p}$ & &$\frac{df}{dx}$  \\
      Transient Jacobian ($\beta=\frac{d\dot{x}}{dx}$) & $x$,$\dot{x}$& ${p}$, $\beta$& &$\beta \frac{df}{d\dot{x}}+\frac{df}{dx}$  \\
      Directional Derivative & $x$,${v}$ & ${p}$& $\frac{df}{dx} \cdot {v}$ & \\
      Sensitivity& $x$ & ${p}$& $\frac{df}{d{p}}$ &  \\
      Hessian$\cdot$Vector& $x$,${v}$ & ${p}$ & & $\frac{d^2f}{dx^2} \cdot {v}$  \\
      Stochastic Galerkin Residual& $x({\xi})$ & ${\xi}$ & $F$ &  \\
      Stochastic Galerkin Jacobian& $x({\xi})$ & ${\xi}$ & & $\frac{dF}{dx}$  \\
      \hline
    \end{tabular}
  \end{center}
\end{table}

It is critical to note that the TBGP machinery is NOT applied to the
linear algebra structures used by the solvers and analysis algorithms.
None of the operator-overloading or expression templating
infrastructure comes into play at this level.  The TBGP is applied
locally on a single element in the assembly process used to fill the
linear algebra data structures.  For example, in our Trilinos-based
implementations the vectors and matrices are objects from the Epetra
or Tpetra libraries.  These are convenient because of their built-in
support for distributed-memory parallelism and their compatibility
with all the solvers in Trilinos (both linear solvers and analysis
algorithms).  However, none of the subsequent implementation of the
TBGP code is dependent on this choice.

Inside the PDE assembly for finite element codes, it
is natural to have element-based storage layout. All of
the computations of the discretized PDE equations operate
on multi-dimensional arrays (MDArrays) of data which can be accesses
with local element-level indexing (local nodes, local
quadrature points, local equation number, etc.).  The current MDArray domain model is specified and implemented in the Shards package in Trilinos \cite{ShardsWebSite}.

While the TBGP computations occur locally within an element, the
assembly of element contributions to the linear algebra objects is
done on local blocks of elements called ``worksets''.  A workset is a
homogeneous set of elements that share the same local bookkeeping and material
information.  While all the computations within each element in a
workset are independent, the ability to loop over a workset amortizes
the overhead of function calls and gives flexibility to obtain
speedups through vectorization, cache utilization, threading, and
compute-node based parallelism. By restricting a workset of elements
to be homogeneous, we can avoid excessive conditional (``if'') tests
or indirect addressing within the workset loops.  The number of
elements in a workset, $N_e$, can be chosen based on a number of
criteria including runtime performance optimization or memory
limitations.

The other dimensions of the MDArrays can include number of local nodes
$N_n$, number of quadrature points $N_q$, number of local equations or
unknowns $N_{eq}$, and number of spatial dimensions $N_d$.  For
instance, a nodal basis function MDArray has dimensions $[N_e, N_n,
N_q]$, while the gradient of the solution vector evaluated at
quadrature points is dimensioned $[N_e, N_q, N_{eq}, N_d]$.

All of the MDArrays are templated on the Scalar data type,
called \texttt{ScalarT} in our code examples.
Depending on what specific Scalar type they are instantiated with,
they will not only hold the value, but can also hold other information
such as the
derivatives (in the case of Jacobian evaluations), sensitivities,
or polynomial chaos coefficients.

At this point, we hope the reader has an understanding of the
template-based generic programming approach from the previous paper,
with the seed-extract-compute paradigm. This current Section \ref{sect:fem}
has
motivated the application of the TBGP approach at the local
or element level, and has defined the distinction between
global linear algebra objects (matrices and vectors that span
the mesh and typically are of double data type) and MDArrays
(element-based data structures of quantities that are
templated on the Scalar type). With this foundation, the
main concept in this paper can be now presented in the
following Section \ref{sect:pde}.

\section{Template Based Element Assembly} \label{sect:pde}

The template-based generic programming approach requires
a {\em seed} phase where the Scalar data types are initialized
appropriately.  As described above in Section \ref{sect:fem},
there are different data structures that need to exist in
the solution phase from the assembly phase. Notably, a {\em gather}
routine is needed to pull in global information (such as the solution
vector from the nonlinear solver) to the local element
data structures (such as the solution values at local nodes in an element).
In our design, we perform the {\em gather} and {\em seed}
operations in the same routine. When we pull global data
into a local data storage, we not only copy it into local
storage, but also seed the Scalar data types as needed.
The seeding is dependent on the Scalar data type, so the
gather operation must be template specialized code. For example,
for a Jacobian evaluation, the partial derivative array associated
with the solution vector is seeded with the identity matrix.

The inverse is also true. At the end of the PDE assembly there
are element-based contributions to the global residual vector
and, depending on the Scalar type, information for the
Jacobian, polynomial chaos expansion, or other evaluation
types contained in the data structure as well. These quantities
need to be {\em extracted} from these data structures as
well as {\em scattered} back from the local to global data
storage containers. We combine the {\em scatter} and {\em extract}
operations into a single step, which again require
template-specialized code. For the Jacobian example, the
derivative array associated with the residual entries
are rows of the element stiffness matrix.

The {\em compute} phase operates solely on local MDArray
data structures with data templated on the Scalar type.
This phase can be written entirely on the generic
template type.  

\begin{figure}[h] \begin{center}
\includegraphics[width=4.5in]{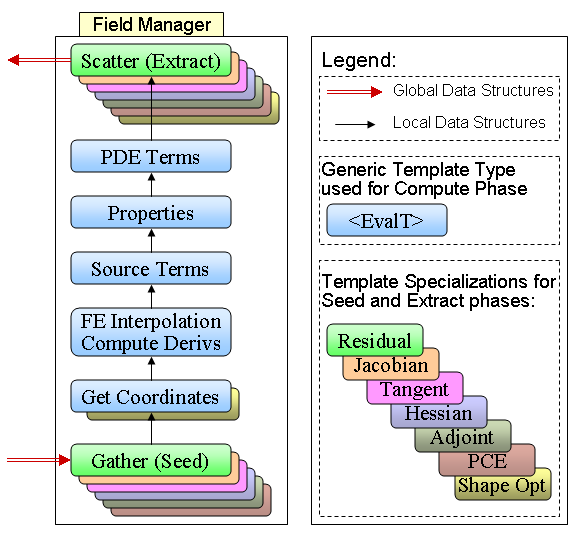}
\caption{A schematic of the template based generic programming
model for PDEs. The Gather (Seed) evaluator takes global data and
copies it into local storage, seeding any embedded data
types in template specialized codes. The Scatter (extract)
phase does the inverse. The Compute phase can be written
on the generic template type, operating on local data.}
\label{fig:rainbow} \end{center} \end{figure}

We have attempted to capture this concept schematically
in Figure \ref{fig:rainbow}. The Gather/Seed phase must
take global data and, depending on the template type,
seed the local arrays appropriately. The compute phase,
broken into five distinct evaluations in this cartoon
(the blue boxes), performs the element level finite
element calculations for the specific PDEs, and is written
just on the generic template type. (The uniqueness of the
Gather Coordinates box will be addressed later in Section \ref{sect:coords}.)
The Scatter/Extract phase takes the results of the assembly and loads the
data into the appropriate global quantities, as dictated
by the specific evaluation type.

The execution of the phases is initiated by the application code and
handled by the Phalanx package \cite{PhalanxWebSite} by traversing the
evaluation kernels in the directed acyclic graph. More details on the
three phases for a typical finite element assembly will be described
in the subsequent sections.

\subsection{Finite Element Assembly Process under TBGP}
Framing the finite element assembly process in terms of the
template-based generic programming concept is best explained by
example.  Here we apply the Galerkin finite element method to a
generic scalar multidimensional conservation equation (see, for example,
\cite{FEFlow2003})
\begin{equation}
\label{eqn:conservation_pde}
\dot{u} + \nabla \cdot \mathbf{Q} + s = 0,
\end{equation}
where $u$ is the unknown being solved for (i.e. the degree of
freedom) and {$\dot{u}$ its time derivative. The flux $\mathbf{Q}$ and source term $s$
are functions of $u$, time, and position.  While the exact form of the flux is not important, we comment that if the flux is strongly convecting, then additonal terms such as SUPG \cite{HughesBrooks79,BrooksHughes82a} may be required to damp non-physical oscillations.  To simplify the analysis, we ignore such terms here.  This is valid for systems where convection is not dominant such as low Reynolds number flows or heat conduction in a solid.
Equation~\eqref{eqn:conservation_pde} is put into variational form which, after
integration by parts and ignoring boundary contributions for the sake
of simplicity, yields the residual equations
\begin{equation}
\label{eqn:residual}
R_u^i = \int_\Vol \left[ \phi^i \dot{u} - \grad\phi^i \bcdot \vector{Q} + \phi^i s \right] \dV.
\end{equation}
In~\eqref{eqn:residual}, $\Vol$ is the domain over which the problem
is solved and $\phi^i$ are the finite element basis functions.  The unknown, its time derivative,  and its spatial derivative are computed using
\begin{equation}
u = \sum_{i=1}^{N_u} \phi^i u^i, \dot{u} = \sum_{i=1}^{N_u} \phi^i \dot{u}^i
\label{eq:fem}
\end{equation}
\begin{equation}
\frac{\partial u}{\partial x_j} = \sum_{i=1}^{N_u} \frac{\partial \phi^i}{\partial x_j} u^i,
\end{equation}
where $u^i$ are the unknown coefficients of the discretization of $u$, $x_j$ is the coordinate direction, and
$N_u$ is the number of basis functions.  The integrations
in~\eqref{eqn:residual} are performed using numerical quadrature,
\begin{equation}
\label{eqn:discrete}
\hat{R}_u^i = \sum_{e=1}^{N_E}\sum_{q=1}^{N_q} \left[\phi^i \dot{u} - \grad\phi^i\bcdot\vector{Q} + \phi^i s\right] w_q |j| = 0,
\end{equation}
where $N_E$ is the total number of elements in the domain, $N_q$ is the
number of quadrature points in an element for the integration order, $|j|$ is the
determinant of the Jacobian of transformation from the physical space
to the element reference space, and $w_q$ the quadrature weights.

With the finite element assembly algorithm defined
by~\eqref{eqn:discrete}, the process can be redefined in terms of the
gather-compute-scatter operations.  The assembly algorithm
in~\eqref{eqn:discrete} loops over the elements in the domain and sums
the partial contributions to form the residual equations,
$\hat{R}_u^i$.  The complete set of residual equations constitute the
global residual, $f$.  Reformulating in terms of the workset concept,
the assembly process for evaluating a residual is defined as,
\begin{equation}\label{eq:f-element-sum}
f(x) = \sum_{k=1}^{N_w} f_k = \sum_{k=1}^{N_w} S_k^T \bar{R}_{u_k}^i (G_k x).
\end{equation}
Here $N_w$ is the number of worksets and $f_k$ is the partial residual
associated with the finite element contributions for the elements in
workset $k$.  $G_k$ is the {\it{gather}} operation that maps the
global solution vector, $x$, to the local solution vector for workset
$k$.  As mentioned above, in the software implementation, the gather
routine also performs the seeding of scalar types.  $S_k^T$ is the
{\it{scatter}} operation that maps the local element residual for
elements of workset $k$ into the global residual contribution, $f_k =
S_k^T \bar{R}_{u_k}^i$. As noted above, in the software
implementation, the extraction process occurs during the scatter.
$\bar{R}_{u_k}^i$ are the element residual contributions to $\hat{R}_u^i$
that come from the
elements in workset $k$ as a function of the local workset solution
vector $G_k x$,
\begin{equation}
\bar{R}_{u_k}^i = \sum_{e=1}^{N_e}\sum_{q=1}^{N_q} \left[\phi^i \dot{u} - \grad\phi^i\bcdot\vector{Q} + \phi^i s\right] w_q |j| = 0.
\end{equation}
$N_e$ is the number of elements in workset $k$.

The important point to note is that while all of the code written
above was used for evaluating a residual, the bulk of the code can now
be reused for other evaluation types such as Jacobians, parameter
sensitivities, stochastic residuals, etc.  This is accomplished merely
by writing an additional specialization for the gather, $G_k$, and
scatter, $S_k^T$, operations {\it{only}}.  All of the code for the
residual evaluation, $\bar{R}_{u_k}^i$, is written {\it{once}} for a
generic template argument for the scalar type and is reused for each
evaluation type.

In the following sections, we now show examples and further explain
each of the assembly steps.

\subsection{Seed \& Gather Phase: Template Specialization} \label{sec:seed}

In this first phase, the approach is to do the gather operation,
$G_k$, (pulling quantities from a global vector) and the seed phase
(initializing the template type for the desired embedded operation) in
the same block of code. In this example, which is an adaptation of
working code, the Phalanx evaluator called
{\texttt{GatherSolution}} is where this operation occurs, and within
the {\texttt{evaluateFields}} method in particular.  As described in the previous paper~\cite{PPS:2011}, the
Trilinos package Phalanx~\cite{PhalanxWebSite} is used to build the governing equations, where
separate pieces of the computation are broken into Phalanx evaluator
objects.

The field to be evaluated is called \texttt{local\_x}, the solution
vector at the local nodes of each element. It depends on
\texttt{global\_x}, which is the solution vector in the vector data
layout. In Figure~\ref{fig:seedResid},
\begin{figure}[tb]
\begin{verbatim}
void GatherSolution<EvaluationType::Residual>::
evaluateFields() {
  // NOTE:: local_x is a 2D MDArray of dimension (numberOfElements,numberOfLocalNodes)
  //        of data type:  double
  for (int elem=0; elem < numberOfElements; elem++)
    for (int node=0; node < numberOfLocalNodes; node++)
      local_x(elem,node) = global_x(ConnectivityMap(elem,node));
}
\end{verbatim}
\caption{Seed and Gather code for Residual evaluation. The \texttt{ConnectivityMap}
function is the degree of freedom, or connectivity map, that gets
the global ID from the element number and local node number. The
Seed phase is trivial, just a copy of the value.}
\label{fig:seedResid}
\end{figure}
 the {\texttt{GatherSolution}}
class is specialized to the {\texttt{Residual}} evaluation type.
This routine simply copies the values from one data structure
to another with the use of a bookkeeping function \texttt{ConnectivityMap}.
In Figure~\ref{fig:seedJac},
\begin{figure}[tb]
\begin{verbatim}
void GatherSolution<EvaluationType::Jacobian>::
evaluateFields() {
  // NOTE:: local_x is a 2D MDArray of dimension (numberOfElements,numberOfLocalNodes)
  //        of data type:  Sacado::FAD with allocated space for
  //                       numberOfLocalNodes partial derivatives.
  for (int elem=0; elem < numberOfElements; elem++) {
    for (int node=0; node < numberOfLocalNodes; node++) {
      local_x(elem,node).val() = global_x(ConnectivityMap(elem,node));
      // Loop over all nodes again, and Seed  dx/dx=I
      for (int wrt_node=0; wrt_node < numberOfLocalNodes; wrt_node++) {
        if (node == wrt_node) local_x(elem,node).dx(wrt_node) = 1.0;
        else                  local_x(elem,node).dx(wrt_node) = 0.0;
      }
    }
  }
}
\end{verbatim}
\caption{Seed and Gather Code for Jacobian evaluation. The Gather operation
is the same as the Residual calculation. The Seed phase involves \texttt{local\_x},
which is now an automatic differentiation data type. The method \texttt{local\_x.val()} accesses
the value, and \texttt{local\_x.dx($i$)} accessing the $i^{th}$ partial derivative. The example
here assumes one equation and one unknown per local node.}
\label{fig:seedJac}
\end{figure}
the code for the {\texttt{GatherSolution}}
class specialized to the {\texttt{Jacobian}} evaluation type is shown.
In addition to the gather operation to load the value of
$x$ into the local data structure, there is also a
seed phase to initialize the partial derivatives with the identity matrix. Here, the independent variables
are defined by initializing the partial derivative array of
the Sacado::FAD automatic differentiation data type. Two nested
loops over the local nodes are used to set $\frac{dx_i}{dx_j}$
to $1.0$ when $i=j$ and to $0.0$ otherwise.

As the number of output quantities to be produced by the
finite element assembly increases (such as those defined
by the rows in Table~\ref{tab:embedded}), so does
the number of template specialized implementations of the
\texttt{GatherSolution} object need to be written. The syntax here is
dependent on the implementation in the Sacado package in Trilinos,
but the concept of seeding automatic differentiation calculations is general.

\subsection{Compute Phase: Generic Template} \label{sec:compute} The
compute phase that computes the local contributions,
$\bar{R}_{u_k}^i$, for a PDE application operates on data that exists
in the local element-based data structures. This code is written
entirely on the generic evaluation template type \texttt{EvalT}.  One
must just write the code needed to evaluate the residual equation, but
using the \texttt{ScalarT} data type corresponding to the evaluation
type \texttt{EvalT} instead of raw double data type (as shown in
Section 7.1 of~\cite{PPS:2011}). The overloaded data type, together
with specializations in the Seed and Extract phases, enable the same
code to compute all manner of quantities such as the outputs in
Table~\ref{tab:embedded}.

In this section we give two examples of the \texttt{evaluateFields}
method of a Phalanx evaluator class. The first is shown in Figure~\ref{fig:computeSource}
which calculates a source term in a heat equation,
\begin{equation}
s = \alpha + \beta u^2,
\end{equation}
where $s$ is the source term from Equation \ref{eqn:conservation_pde},
$\alpha$ and $\beta$ are parameters in this model, and $u$ is the solution field.
The code presupposes that the field $u$
has been computed in another evaluator and is an MDArray
over elements and quadrature points.  The factors $\alpha$ and $\beta$ in this
example are scalar values that do not vary over the domain. (We will
discuss later in Section~\ref{sect:params} how to expose $\alpha$ and $\beta$ as
parameters for design or analysis.)  Note again that this code is
templated on the generic \texttt{EvalT} evaluation type, and only one
implementation is needed.

\begin{figure}
\begin{verbatim}
void SourceTerm<EvalT>::
evaluateFields() {
  // NOTE:: This evaluator depends on properties "alpha" and "beta"
  //        and scalar quantity "u" to compute the scalar source "s"
  for (int elem=0; elem < numberOfElements; elem++) {
    for (int qp=0; qp < numberOfQuadPoints; qp++) {
      source(elem,qp) += alpha + beta * u(elem,qp) * u(elem,qp);
    }
  }
}
\end{verbatim}
\caption{Example of an evaluation kernel from the Compute phase. Note that
the code is templated on the Generic EvalT evaluation type. This code will
propagate the auxiliary information contained in \texttt{EvalT::ScalarT}
data type for any
of the embedded capabilities. No template specialization is needed.}
\label{fig:computeSource}
\end{figure}

For general PDE codes, it is common, and efficient, to use discretization libraries
to perform common operations. For a finite element code, this
includes basis function calculations, calculating the transformation
between reference and physical elements, and supplying quadrature
schemes. To the extent that these operations occur within the
seed-compute-extract loop, they must support templating for the TBGP approach
to work. In Trilinos, the Intrepid
discretization library \cite{Intrepidspci} serves all these roles, and was
written to be templated on the generic \texttt{ScalarT} data type.

Figure~\ref{fig:Heat} shows an example evaluator method for the final
assembly of the residual equation for a heat balance.  This code uses
the \texttt{Integrate} method of the Intrepid finite element library
to accumulate the summations over quadrature points of each of the
four terms. The integrals are from the variational formulation of the
PDEs, where the terms are matched with the Galerkin basis functions
(adapted from equation~\eqref{eqn:residual} but using common notation
for heat transfer):
\begin{equation}
f = \int_{\Vol}{ ( \mathbf{Q} \cdot \nabla \phi^i ) \dV} + \int_{\Vol}{ (s \phi^i ) \dV}+ \int_{\Vol}{ (\dot{T} \phi^i ) \dV}
\label{eq:fheat}
\end{equation}
where $\phi^i$ are the basis functions. The three terms on the right
hand side correspond to the diffusive, source, and accumulation terms.

The source term $s$ in this equation can be a function of
the solution, a function of the position, or a simple constant.
The dependencies must be defined in the source term evaluator.
However, these dependencies do not need to be described in
the heat equation residual. This same piece of code will
accurately propagate any derivatives that were seeded in
the gather phase, and accumulated in the source term evaluator.
\begin{figure}
\begin{verbatim}
void HeatEquationResidual<EvalT>::
evaluateFields() {
  // NOTE:: This evaluator depends on several precomputed fields,
  //        flux, source, Tdot (time derivative),
  //        wBF (Basis Function with quadrature and transformation weights)
  //        and wGradBF (gradient of Basis Functions with weights).
  //        The result is the TResidual field, the element contribution
  //        to the heat equation residual.

  typedef Intrepid::FunctionSpaceTools FST;

  FST::integrate<ScalarT>(TResidual, flux, wGradBF);

  FST::integrate<ScalarT>(TResidual, source, wBF);

  FST::integrate<ScalarT>(TResidual, Tdot, wBF);
}
\end{verbatim}
\caption{Evaluator for the final assembly of the heat equations. The
terms correspond to those in Eq.~\ref{eq:fheat}, in order. The variable
\texttt{TResidual} is being accumulated in each step. All variables
are MDArrays. Depending on the template parameter \texttt{EvalT} and
the corresponding data type \texttt{ScalarT}, this same block of code
is used for accumulating the residual, Jacobian, or any of the output
quantities listed in Table~\ref{tab:embedded}.}
\label{fig:Heat}
\end{figure}

We should also note that it is possible to write template-specialized
code for the compute phase. If, for instance, one would like to
hand-code the Jacobian fill for efficiency, or to leave out terms
for a preconditioner, once could simply write a function:
\begin{verbatim}
HeatEquationResidual<EvaluationType::Jacobian>::evaluateFields()
\end{verbatim}
where
the generic template type \texttt{EvalT} is replaced by the
template specialized evaluation type of \texttt{Jacobian}.

\subsection{Extract \& Scatter Phase: Template Specialization} \label{sec:extract}

This section closely mimics Section~\ref{sec:seed}, but with
the transpose operations. Here, each local element's contributions to
the finite element residual is scattered into the global data
structure. At the same time, if additional information is stored
in the templated data types, it is extracted and scattered into the global linear algebra objects.

In this example, the evaluator object called
{\texttt{ScatterResidual}} is where this operation occurs, and within
the {\texttt{evaluateFields}} method in particular. The culmination of
all the compute steps above have resulted in the computation
of the field \texttt{local\_f}, which represents the local element's
contribution, $\bar{R}_{u_k}^i$, to the global residual vector, but may also
contain additional information in the \texttt{ScalarT} data type.

In Figure~\ref{fig:extractResid}, the {\texttt{ScatterResidual}}
class is specialized to the {\texttt{Residual}} evaluation type.
This routine simply copies the values from the element data structure
to the global data structure with the use of the same
bookkeeping function \texttt{ConnectivityMap} that was used in Section~\ref{sec:seed}.
In Figure~\ref{fig:extractJac}, the code for the {\texttt{ScatterResidual}}
class specialized to the {\texttt{Jacobian}} evaluation type is shown.
Here, the local, dense, stiffness matrix is extracted from
the Sacado::FAD automatic differentiation data type. Two nested
loops over the local nodes are used to extract $\frac{df_i}{dx_j}$
and load them into the global sparse matrix.

\begin{figure}
\begin{verbatim}
void ScatterResidual<EvaluationType::Residual>::
evaluateFields() {
  // NOTE:: local_f is a 2D MDArray of dimension (numberOfElements,numberOfLocalNodes)
  //        of data type:  double
  for (int elem=0; elem < numberOfElements; elem++)
    for (int node=0; node < numberOfLocalNodes; node++)
      global_f(ConnectivityMap(elem,node)) = local_f(elem,node);
}
\end{verbatim}
\caption{Extract and Scatter Code for Residual evaluation. The ConnectivityMap
function is also called the degree of freedom map, and gets
the global ID from the element number and local node number. The
Extract phase is trivial for this Evaluation type, just a copy of the value.}
\label{fig:extractResid}
\end{figure}

\begin{figure}
\begin{verbatim}
void ScatterResidual<EvaluationType::Jacobian>::
evaluateFields() {
  // NOTE:: local_f is a 2D MDArray of dimension (numberOfElements,numberOfLocalNodes)
  //        of data type:  Sacado::FAD with allocated space for
  //                       numberOfLocalNodes partial derivatives.
  for (int elem=0; elem < numberOfElements; elem++) {
    for (int node=0; node < numberOfLocalNodes; node++) {
      int row=ConnectivityMap(elem,node);
      for (int wrt_node=0; wrt_node < numberOfLocalNodes; wrt_node++) {
        int col=ConnectivityMap(elem,wrt_node);
        double val = local_f(elem,node).dx(wrt_node); //Extract df_i/dx_j
        AddSparseMatrixEntry(row,col,val);
      }
    }
  }
}
\end{verbatim}
\caption{Extract and Scatter Code for Jacobian evaluation.
The Extract phase involves local\_f,
which is now an Automatic Differentiation data type. The method dx($i$) on this
data type accesses the $i^{th}$ partial derivative. A (fictitious) method called
AddSparseMatrixEntry(int row, int col, int value) shows how this Jacobian
information is scattered into the global and sparse $\frac{df}{dx}$ storage.}
\label{fig:extractJac}
\end{figure}

A set of template specialized implementations of the
{\texttt{ScatterResidual}} object need to be written to match those in the
{\texttt{GatherSolution}} class. Just two are shown here. These
implementations are specific to the interface to the global
data structures being used, which here are encapsulated
in the {\texttt{ConnectivityMap()}} and {\texttt{AddSparseMatrixEntry()}} methods.

A central point to this paper, and the concept of template-based generic programming,
is that the implementations in the Seed \& Gather and the
Extract \& Scatter sections can be written agnostic to
the physics being solved. While the work of correctly programming these
two phases for all evaluation types is not at all trivial,
the development effort is completely {\em orthogonal} to the work
of adding terms to PDEs. Once a code
is set up with implementations for a new evaluation type,
it is there for any PDEs assembled in the compute phase.  

We note that while the examples shown here are trivial, the design of
the assembly engine is very general and allows for complex
multiphsyics problems.  In particular, unknowns are not all bound to
the same basis.  Fully-coupled mixed basis problems have been
demonstrated with the Phalanx assembly engine.  The design of each
evaluator in the graph is completely controlled by the user, thus
allowing for any algorithm local to the workset to be implemented.

We further note that for all of the evaluation routines above, the loops were explicitly written in the evaluator.  In general, this is not ideal since it is likely to be repeated across multiple evaluators and can introduce additional points for error.  These loops could be eliminated using utility functions or expression templates \cite{VelhuizenET}.  This will be a future area of research.  Furthermore, optimizing the ordering of the nested loops and the corresponding data layouts of the field data and embedded scalar types are also areas of future research.

Finally, we note that during the design of the Phalanx package, great care was given to designing the library so that users with little C++ template experiience could easily add new physics.  We feel this has been extremely successful in that there exists over 10 distinct physics applications using the TBGP packages in Trilinos.  The drawback, however, is that the initial setup of the TBGP process requires a programmer with a strong background with templates.  In contrast, DSL-based codes such as Sundance \cite{Long2010} and FEniCS \cite{FenicsOverview2007,LoggMardalEtAl2012a} automate the entire assembly process for users lowering the barrier for adoption.  We feel that the extra work in setting up the TBGP machinery is worth the effort as we have very quickly extended the embedded analysis support to new types such as the stochastic galerkin methods.

\section{Extensions to TBGP for Finite Element Code Design}\label{sect:pdefill}

The basic implementation details for template-based generic programming
approach for finite element code were described in the previous section.
As we have implemented this approach in application codes, we have run
across many issues, and implemented solutions to them. Some of the most
important of these are described in the following sections.

\subsection{Infrastructure for Exposing Parameters/Properties \label{sect:params}}

One of the main selling points of the template-based generic
programming approach is the ability to perform design and analysis
involving system parameters. Continuation, sensitivity analysis,
optimization, and UQ all require that model parameters be manipulated
by the analysis algorithms. These parameters are model specific,
and commonly include a value of a boundary condition, a dimensionless
group such as the Reynolds number, a model parameter such as an Arrhenius
rate coefficient, or a shape parameter such as the radius of some
cylindrical part. In this section we briefly describe our infrastructure
for exposing parameters.

Infrastructure for dealing with parameters should try to meet
the following design requirements: a simple interface for model
developers to expose new parameters, integration into the template-base
approach so derivatives with respect to parameters are
captured, and seamless exposure of the parameters to design
algorithms such as optimization and UQ.

The approach that has been successful in our codes has been to
use the \texttt{ParameterLibrary} class in the Sacado package of Trilinos. This
utility stores the available parameters by string name and
value, and can handle the multiple data types needed by the
template-based approach. The developer can register parameters
in the parameter library, identified by strings, by simply calling the register method
during the construction phase. To expose the $\alpha$ and
$\beta$ parameters in the above Example~\ref{fig:computeSource}
by labels ``Alpha'' and ``Beta'',
the constructor for the \texttt{SourceTerm} evaluator simply needs to
add the lines:
\begin{verbatim}
parameterLibrary.registerParameter("Alpha",this);
parameterLibrary.registerParameter("Beta",this);
\end{verbatim}
assuming that a \texttt{parameterLibrary} object is in scope.
At the end of the problem construction, the parameter library can be
queried for a list of registered parameters, and will include
these two in addition to those registered elsewhere.

The analysis
algorithms can then manipulate the values of these parameters in
the \texttt{ParameterLibrary}. There is a choice of using a push or a pull
paradigm: when the value is changed in the parameter library, is
it immediately pushed to the evaluator where it will eventually
be used, or is it up to the model to pull the parameter values from
the parameter library when needed. We have chosen the push approach,
since with this choice there is no performance penalty for exposing
numerous parameters as potential design variables. Parameters are only
pushed to one location, so parameters that are used in multiple evaluators
must have a root evaluator where they are registered and other evaluators
must have a dependency on that one.

Any evaluator class that registers a parameter must inherit from
an abstract \texttt{ParameterAccessor} class, which has a single method
called  \texttt{ ScalarT\& getValue(std::string name)}.
Any parameter that gets registered with the \texttt{ParameterLibrary} needs
to send a pointer to a \texttt{ParameterAccessor} class, so the parameter library
can push new values of the parameter when manipulated by an analysis
algorithm. In the example above, this is handled by the \texttt{this}
argument in the registration call.

For this example, the \texttt{getValue} method can be simply implemented
as shown in Figure~\ref{fig:get_value}, assuming parameters
\texttt{alpha} and \texttt{beta} are member data
of generic template type \texttt{ScalarT}. Sensitivities of the
residual equation with respect to parameters are calculated
with automatic differentiation when evaluated with the
\texttt{Tangent} evaluation type. Like the \texttt{Jacobian} evaluation
type, the associated Scalar data type is a Sacado::FAD type.
However, the length of the derivative array is the number of
parameters, and the seed and extract phases require different
specializations.

\begin{figure}
\begin{verbatim}
ScalarT& SourceTerm<EvalT>::getValue(std::string name)
{
  if (name=="Alpha") return alpha;
  elseif (name=="Beta") return beta;
}
\end{verbatim}
\caption{Example implementation of \texttt{getValue} method, which provides a
hook for analysis algorithms to manipulate design parameters.} \label{fig:get_value}
\end{figure}

\subsection{Shape Optimization: A Second Scalar Type}

Quantities in the PDE assembly that might have nonzero
partial derivatives with respect to an independent variable
(whether it be a parameter or part of the solution vector)
must be a have a templated data type. In this way, derivatives
can be propagated using the object overloading approach.
Constants (such as $\pi$) can be hardwired to the
\texttt{RealType} data type to avoid the expense of propagating
partial derivatives that we know are zero.

For the bulk of our calculations, the coordinates of the nodes in
our finite element mesh are fixed. All the quantities that are
solely a function of the coordinates, such as the basis function
gradients and the mapping from an element to the reference element,
can be set to \texttt{RealType}. However, when we began to
do shape optimization, the coordinates of the node could now have
nonzero derivatives with respect to the shape parameter in the
\texttt{Tangent} (sensitivity) evaluation. To simply make
all quantities that are dependent on the coordinates to
be the \texttt{ScalarT} type would trigger an excessive amount of
computations, particularly for the Jacobian calculations,
where the chain rule would be propagating zeroes through
a large part of the finite element assembly.

The solution was to create a second generic data type,
\texttt{MeshScalarT}, for all the quantities that have
non-zero derivatives with respect to the coordinates,
but have no dependency on the solution vector. The
Traits class defined in the previous paper is extended to include
\texttt{MeshScalarT} as well as \texttt{ScalarT} as follows.
\begin{verbatim}
  struct UserTraits : public PHX::TraitsBase {

    // Scalar Types
    typedef double RealType;
    typedef Sacado::FAD FadType;

    // Evaluation Types with default scalar type
    struct Residual { typedef RealType ScalarT; typedef RealType MeshScalarT;};
    struct Jacobian { typedef FadType ScalarT;  typedef RealType MeshScalarT;};
    struct Tangent { typedef FadType ScalarT;  typedef FadType MeshScalarT;};
    .
    .
    .
  }
\end{verbatim}

If, in the future we decide to do a moving mesh problem, such as thermo-elasticity,
where the coordinate vector for the heat equation does depend
on the current displacement field as calculated in the elasticity
equation, then the \texttt{Jacobian} evaluation could be switched
in this traits class to have \texttt{typedef FadType MeshScalarT}.
Automatically, the code would calculate the accurate Jacobian
for the fully coupled moving mesh formulation.

While it can be complicated to pick the correct data type for
all quantities with this approach, the Sacado implementation of
these data types has a useful feature. The code will not compile
if you attempt to assign a derivative data type to a real type.
The casting away of derivative information must be done explicitly,
and can not be done by accident. This is illustrated in the
following code fragment, as annotated by the comments.
\begin{verbatim}
RealType r=1.0;
FadType f=2.5;
f = r;  //Allowed. All derivatives set to zero.
r = f;  //Compiler will report an error.
\end{verbatim}

\subsection{Template Infrastructure}\label{sect:template}
Template-based generic programming places a number of additional
requirements on the code base. Building and manipulating the objects
can be intrusive.  Additionally the compile times can be excessive as
more and more template types are added to the infrastructure. Here we
address these issues.

\subsubsection{Extensible Infrastructure}
The infrastructure for a template-based assembly process must be
designed for extensibility.  The addition of new evaluation types
and/or scalar types should be minimally invasive to the code.  To
support this requirement, a template manager class has been developed
to automate the construction and manipulation of a templated class
given a list of template types. 

A template manager instantiates a particular class for each template
type using a user supplied factory for the class.  The instantiated
objects are stored in a {\texttt{std::vector}} inside the manager.  To
be stored as a vector, the class being instantiated must inherit from
a non-templated base class.  Once the objects are instantiated, the
template manager provides functionality similar to a
{\texttt{std::vector}}.  It can return iterators to the base class
objects.  It allows for random access based on a template type using
templated accessor methods, returning an object of either the base or
derived class.  

The list of types a template manager must build is fixed at compile
time through the use of template metaprogramming techniques
\cite{BoostMPLBook2004} implemented in the Boost MPL library
\cite{BoostLib} and Sacado \cite{SacadoWebSite}.  These types are
defined in a ``traits'' class.  This object is discussed in
detail in \cite{PPS:2011}.

We note that the template manager described here is a simplified
version of the tuple manipulation tools supplied by the Boost Fusion
library \cite{BoostLib}.  In the future, we plan to transition our
code to using the Fusion library.

\subsubsection{Compile Time Efficiency}
For each class templated on an evaluation type, the compiler must
build the object code for each of the template types.  This can result
in extremely long compile times even for very minor code changes.
For that reason explicit template
instantiation is highly recommended for all classes that are templated
on an evaluation type.  In our experience, not all compilers can
support explicit instantiation.  Therefore both the inclusion model
and explicit template instantiation are supported in our objects (see
Chapter 6 of \cite{CPPTemplatesBook2003} for more details) .

The downside to such a system is that for each class that implements
explicit instantiation, the declaration and definition must be split
into separate header files and a third (file.cpp) file must also be added to the
code base.

\subsection{Incorporating Non-Templated Third-Party Code} \label{sect:nontemplated}

In some situations, third-party code may provide non-template based implementations of some analysis evaluations.  For example, it is straightforward to differentiate some Fortran codes with source transformation tools such as ADIFOR~\cite{Bischof1996A2A} to provide analytic evaluations for first and higher derivatives.  However clearly the resulting derivative code does not use the template-based procedure or the Sacado operator overloading library.  Similarly some third-party libraries have hand-coded derivatives.  In either case, some mechanism is necessary to translate the derivative evaluation governed by Sacado into one that is provided by the third-party library.  Providing such a translation is a relatively straightforward procedure using the template specialization techniques already discussed.  Briefly, a Phalanx evaluator should be written that wraps the third-party code into the Phalanx evaluation hierarchy.  This evaluator can then be specialized for each evaluation type that the third-party library provides a mechanism for evaluating. This specialization extracts the requisite information from the corresponding scalar type (e.g., derivative values) and copies them into whatever data structure is specified by the library for evaluating those quantities.  In some situations, the layout of the data in the given scalar type matches the layout required by the library, in which case a copy is not necessary (e.g., Sacado provides a forward AD data type with a layout that matches the layout required by ADIFOR, in which case a pointer to the derivative values is all that needs to be extracted).  However this is not always the case, so in some situations a copy is necessary.

Such an approach will work for all evaluation types that the third-party library provides some mechanism for evaluating.  However clearly situations can arise where the library provides no mechanism for certain evaluation types.  In this case the specializations must be written in such a way as to generate the required information non-intrusively.  For example, if the library does not provide derivatives, these can be approximated through a finite-differencing scheme.  In a Jacobian evaluation for example, the Jacobian specialization for the evaluator for this library would make several calls to the library for each perturbation of the input data for the library, combine these derivatives with those from the inputs (dependent fields) for the evaluator (using the chain rule) and place them in the derivative arrays corresponding to the outputs (evaluated fields) of the evaluator.  Similarly polynomial chaos expansions of non-templated code can be computed through non-intrusive spectral projection~\cite{Reagan:2003p920}.

\subsection{Mesh Morphing and Importing Coordinate Derivatives} \label{sect:coords}

The shape optimization capability that will be demonstrated in Section~\ref{sect:app}
requires sensitivities of the residual equation with respect to shape
parameters. Our implementation uses an external
library for moving the mesh coordinates as a function
of shape parameters, a.k.a. mesh morphing. This is an active research area,
and a paper has just been prepared detailing six different approaches on a
variety of applications \cite{staten11}.

Briefly, the desired capability is
for the application code to be able to manipulate shape parameters,
such as a length or curvature of part of a solid model, and for the mesh
morphing utility to provide a mesh that conforms to that geometry. To
avoid changes in data structures and discontinuities in an objective function
calculation, it is desirable for the mesh topology, or connectivity, to
stay fixed. The algorithm must find a balance between maintaining good mesh
quality and preserving the grading of the original mesh, such as anisotropy in
the mesh designed to capture
a boundary layer. Large shape changes that require remeshing are beyond the
scope of this work, and would need to be accommodated by remeshing and restarting
the optimization run.

A variety of mesh morphing algorithms have been developed and investigated. At one end
of the spectrum is the smoothing approach, where the surface nodes are
moved to accommodate the new shape parameters and the resulting mesh is
smoothed until the elements regain acceptable quality. At the other end
of the spectrum is the FEMWARP algorithm \cite{shontz}, where a finite
element projection is used to warp the mesh, requiring a global linear
solve to determine the new node locations. In this paper, we have used
a weighted residual method, where the new node coordinates are based
on how boundary nodes in their neighborhood have moved. We chose to
always morph the mesh from the original meshed configuration to the chosen
configuration, even if an intermediate mesh was already computed at
nearby shape parameters, so that the new mesh was uniquely defined by
the shape parameters.

Since the mesh morphing algorithm
is not a local calculation on each element but operates across the entire
mesh at one time, the derivatives cannot be calculated within a Phalanx
evaluator using the template-based approach, nor using the methods described
in Section~\ref{sect:nontemplated}.

Our approach for calculating sensitivities of the residual vector $f$ with respect to
shape parameters $p$ is to use the chain rule,
\begin{equation}
\frac{\partial f}{\partial p} = \frac{\partial f}{\partial X} \frac{\partial X}{\partial p}.
\end{equation}
where $X$ is the coordinate vector. Outside of the seed-compute-extract section
of templated code, we pre-calculate the mesh sensitivities $\frac{\partial X}{\partial p}$
with a finite difference algorithm around the mesh morphing algorithm. This
multi-vector is fed into the residual calculation as global data.

As part of the typical assembly, there is a \texttt{gatherCoordinates} evaluator that
takes the coordinate vector from the mesh database and gathers it into the
local element-based MDArray data structure. For shape sensitivities, the coordinate
vector is a Sacado::FAD data type, and the
gather operation not only imports the values of the coordinates, but also seeds
the derivative components from the pre-calculated $\frac{\partial X}{\partial p}$
vectors. This is shown schematically in Figure~\ref{fig:rainbow}, where
the \texttt{GatherCoordinates} box is shown to have a template specialized version
for shape optimization in addition to a generic implementation for all other
evaluation types. When the rest of the calculation proceeds, the directional
derivative of $\frac{\partial f}{\partial X}$ in the direction of
$\frac{\partial X}{\partial p}$ is computed. The same implementation for
extracting $\frac{\partial f}{\partial p}$ works for this case as when
$p$ is a set of model parameters as described in Section~\ref{sect:params}.

\section{Demonstration: Sliding Electromagnetic Contact} \label{sect:app}

The template-based generic programming approach is demonstrated here on
a prototype 3D PDE application: the sliding electromagnetic contact problem.
The geometry of this problem is shown in Figure \ref{fig:schematic}, where,
for the nominal design, this 2D geometry is simply extruded into the third
dimension.
A slider (light blue) is situated between two conductors (green and yellow)
with some given shapes of the contact pads (thin red and blue regions). When
a potential difference ($\psi$) over the device is prescribed, an electrical current flows
through system in the general direction of the dashed red line. The electrical
current generates a magnetic field that in turn propels the slider forward.
In addition, the current generates heat.

\begin{figure}[h] \begin{center}
\includegraphics[width=2.0in]{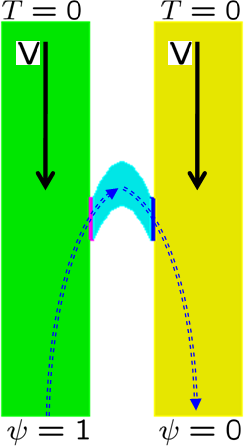}
\caption{A front view of the geometry for the sliding electromagnetic contact
application. The imposed gradient in potential $\psi$ causes electric current,
Joule heating, and a magnetic field that propels the slider (which is not
currently modeled).}
\label{fig:schematic} \end{center} \end{figure}

The design problem to be investigated is to find the shape of the
slider, with a given volume, that minimizes the maximum temperature
achieved inside of it.

\subsection{Governing Equations and Objective Function}

In this demonstration, we simplify the system by decoupling
the magnetics, and solve a
quasi-steady model where the slider velocity $\mbox{v}$ is given.
The model is then reduced to two coupled PDEs. The first is a
potential equation for the electric potential $\psi$. The second
is a heat balance that accounts for conduction, convection,
and Joule heating source term that depends on the current $\nabla \psi$:
\begin{eqnarray}
- \nabla \cdot \sigma \nabla \psi &=& 0,\label{eq:gov1}\\
- \nabla \cdot \kappa \nabla T - {\mbox v} \cdot \nabla T &=& \sigma (\nabla\psi)^2.\label{eq:gov2}
\end{eqnarray}
The electrical conductivity, $\sigma$, varies as a function of the local temperature
field based on a simplified version of Knoepfel's model \cite{knoepfel},
\begin{eqnarray}
\sigma(T) &=& {\sigma_0}/[1+\beta(T-T_0)],
\label{eq:sig}
\end{eqnarray}
where $\sigma_0$ can take different values in the slider, the conductor, 
and in the pads ($\sigma^p_0$).
This dependency results in a coupled pair of equations.

By choosing the frame of reference that stays with the slider, we
impose a fixed convective velocity $\mbox{v}_0$ in the beams, and $\mbox{v}=0$ in
the slider. The Dirichlet boundary conditions for the potential and temperature
are shown in Figure \ref{fig:schematic}, with all others being natural
boundary conditions. Since these equations and
geometry are symmetric about the mid plane (a vertical line in this figure),
we only solve for half of the geometry and impose $\psi=0.5$ along
this axis.

This PDE model was implemented using Phalanx evaluators. By specifying the
dependencies with the evaluators, such as $\sigma(T)$, the evaluation tree
is automatically constructed. The full graph for this problem is shown
in Figure~\ref{fig:graph}. As with the ODE example in the previous paper~\cite{PPS:2011},
only the Gather and Scatter functions need to be written with template
specialization for the Seed and Extract phases. All the intermediate
Compute quantities can be written once on a generic evaluation type.

As an example of how to interpret this graph: the oval marked
$T_q$ computes the temperature field
at the quadrature points using the Basis Functions and $T_n$, the
Temperature field at the nodes (implementing Equation \ref{eq:fem});
$T_q$ is subsequently used to compute
$\kappa$ and $\sigma$ in another evaluator
(which implements Equation \ref{eq:sig}).
\begin{figure}[h] \begin{center}
\includegraphics[width=3.2in]{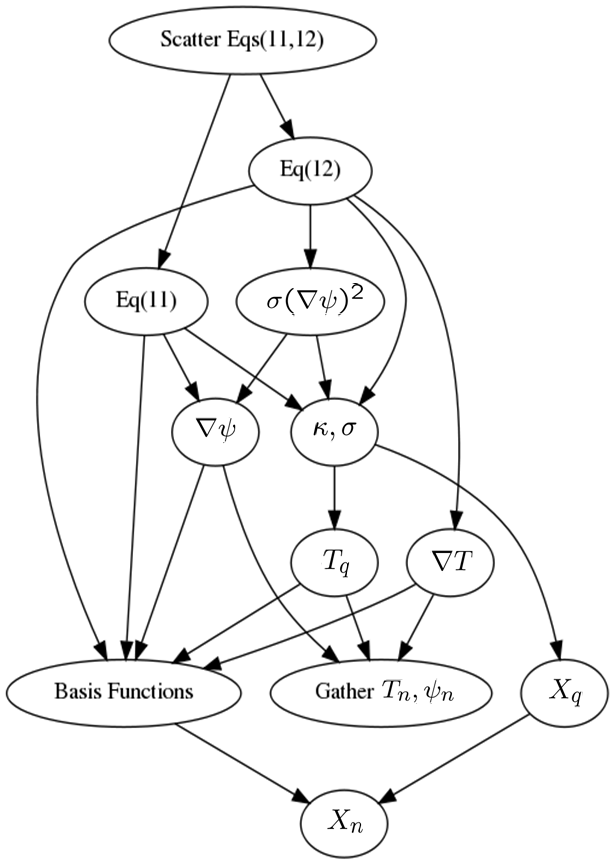}
\caption{The dependency graph for the Phalanx evaluators that build the thermo-electric equation set is shown.
Each box represents a separate class. The quantities at the bottom must be computed before those
above. Derivatives are automatically propagated by the chain rule using the Sacado Automatic Differentiation data types. Here $X$ is the coordinate vector, $T$ the temperature, and $\psi$ the potential, and subscript $n$ and $q$ indicating to node and quadrature point data.}
\label{fig:graph} \end{center} \end{figure}

The objective function $g$, which is to be minimized by the design problem,
is simply
\begin{equation}
g(\psi,T) = \|T\|_{\infty}.
\label{eq:objfn}
\end{equation}

The design parameters $p$ modify the shape of the slider. Its shape is fixed
to match the rectangular pads at each end, and its volume is fixed to
be that of the rectangular box between them. In between, the shape
is allowed to vary parabolically. For a one-parameter optimization problem,
the matched parabolic profiles of the top and bottom of the slider are free
to vary by a single maximum deflection parameter. Because of the symmetry of
the problem, this defines the parabola. For two-parameter optimization
problem, these two parabolas are allowed to vary independently and a third,
the bulge of the slider out of the plane of the figure, is adjusted so that
the volume constraint is met.

\subsection{Gradient-based Optimization}

The optimization problem is solved using a gradient-based optimization
algorithm from the Dakota framework. Dakota can be built as part of Trilinos
using the build system and adaptors in the TriKota package.

The goal is
to minimize the objective function $g(p)$ from Equation \ref{eq:objfn} as a
function of the shape parameters $p$. In addition, the problem is constrained
so that the discretized PDEs are satisfied, $f(x)=0$ where $f$ represents the finite
element residuals for the equation set specified in Equations~\ref{eq:gov1}--\ref{eq:gov2} and the solution vector $x$
is the combined vector of the discretized potential and temperature fields.
The shape parameters do not
appear explicitly in the objective function, or even in the governing equations~(\ref{eq:gov1}--\ref{eq:gov2}), but instead effect the geometry of the problem. They
appear in the discretization, and can be written as $f(x,X(p)))$, where $X$
is the vector of coordinates of nodes in the mesh.

 In addition to the objective function $g(p)$, the gradient-based algorithm
 depends on the reduced gradient of the objective function with respect to
 the parameters. The formula for this term can be expanded as,
\begin{equation}
\frac{dg}{dp} = \frac{\partial g}{\partial p} - {\frac{\partial g}{\partial x}}^T {\frac{\partial f}{\partial x}}^{-1} \frac{\partial f}{\partial X} \frac{dX}{dp}.
\end{equation}
Each of these terms is computed in a different way. Starting at the end, the multi-vector $X_p = \frac{dX}{dp}$ is
computed with finite differences around the mesh morphing algorithm, as described in
Section~\ref{sect:coords}. The sensitivity of the residual vector with respect to the
shape parameters is the directional derivative $\frac{\partial f}{\partial X}$ in the direction $X_p$.
This is computed using Automatic Differentiation using the infrastructure described
in Section~\ref{sect:coords}, where the Sacado automatic derivative data type for the
coordinate vector $X$ is seeded with the derivative vectors $X_p$. The result is
a multi-vector $f_p$.

The Jacobian matrix $\frac{\partial f}{\partial x}$ is computed with automatic differentiation.
All the Sacado data types are allocated with derivative arrays of length $16$,
which is the number of independent variables in a hexahedral element with trilinear
basis functions and two degrees of freedom per node. The local element solution vector
$x$ is seeded with $\frac{\partial x_i}{\partial x_j}=1$ when $i=j$. The action of the inverse of
the Jacobian on $f_p$ is performed with a preconditioned iterative linear solver
using the Belos and Ifpack packages in Trilinos.

The gradient of the objective function $\frac{\partial g}{\partial x}$ is computed by hand,
since $g$ is the max operator on the half of $x$ corresponding to the temperature
unknown. The non-differentiability of the max operator with respect to changes
in parameter can in general be an issue. However, no problems were encountered in this
application since the location of the maximum did nor move significantly over itrations. Finally, the
term $\frac{\partial g}{\partial p}$ was identically zero because the parameters
did not appear explicitly in the objective function.

First, a one-parameter optimization problem was run to find the parabolic
deflection of the top and bottom of the slider that minimized the maximum
temperature. In addition to the gradient-based optimization algorithm
described above, a continuation run was performed using the LOCA
package \cite{locaIJBC} in Trilinos. The results are shown in Figure~\ref{fig:locaDak}.
The continuation run shows the smooth response
surface for a wide range if deflections, both positive and negative.
The optimization iteration rapidly converges to the minimum.
The optimum occurs for a small positive value of the deflection
parameter, corresponding to a shape that is slightly arched upwards but
rather near the nominal shape of a rectangular box.

\begin{figure}[h] \begin{center}
\includegraphics[width=4.5in]{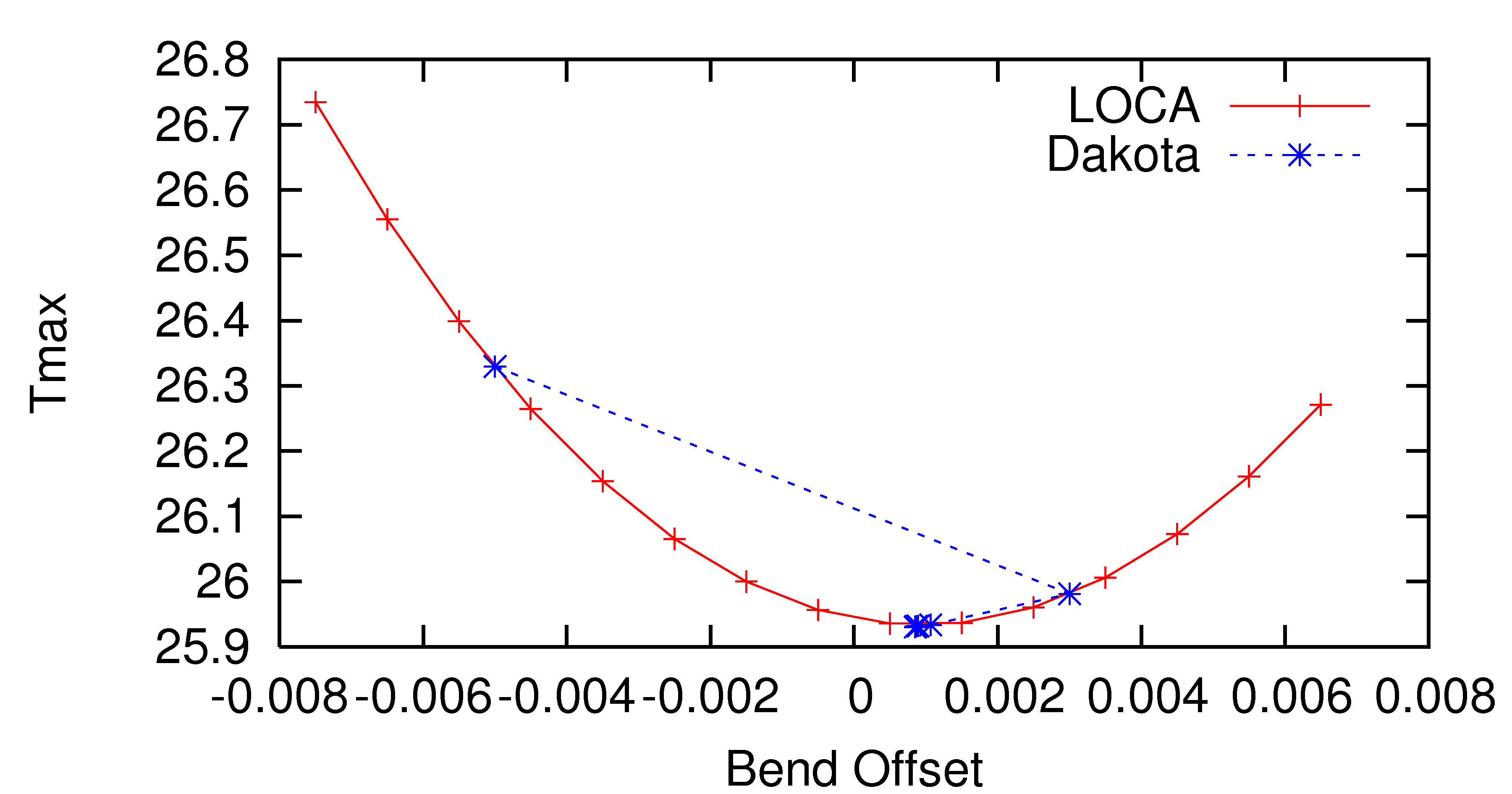}
\caption{Results for a continuation run with LOCA and a minimization
run using Dakota, for a one-parameter shape optimization problem. }
\label{fig:locaDak} \end{center} \end{figure}

A two-parameter optimization run was also performed, where
the top and bottom parabolas were freed to vary independently,
and the bulge of the slider was adjusted to conserve the
volume of the mesh. Figure~\ref{fig:init2} shows the initial
configuration, with a rather arched bottom surface, and nearly
flat upper surfaces, and a moderate bulge. The optimal shape
is shown in Figure~\ref{fig:opt2}. As with the one-parameter case,
the optimal shape was found to be close to a rectangular box.
The temperature contouring of the two figures, which share a
color map, shows a noticeable reduction in the temperature at
the optimal shape.

\begin{figure}[h] \begin{center}
\includegraphics[width=4.2in]{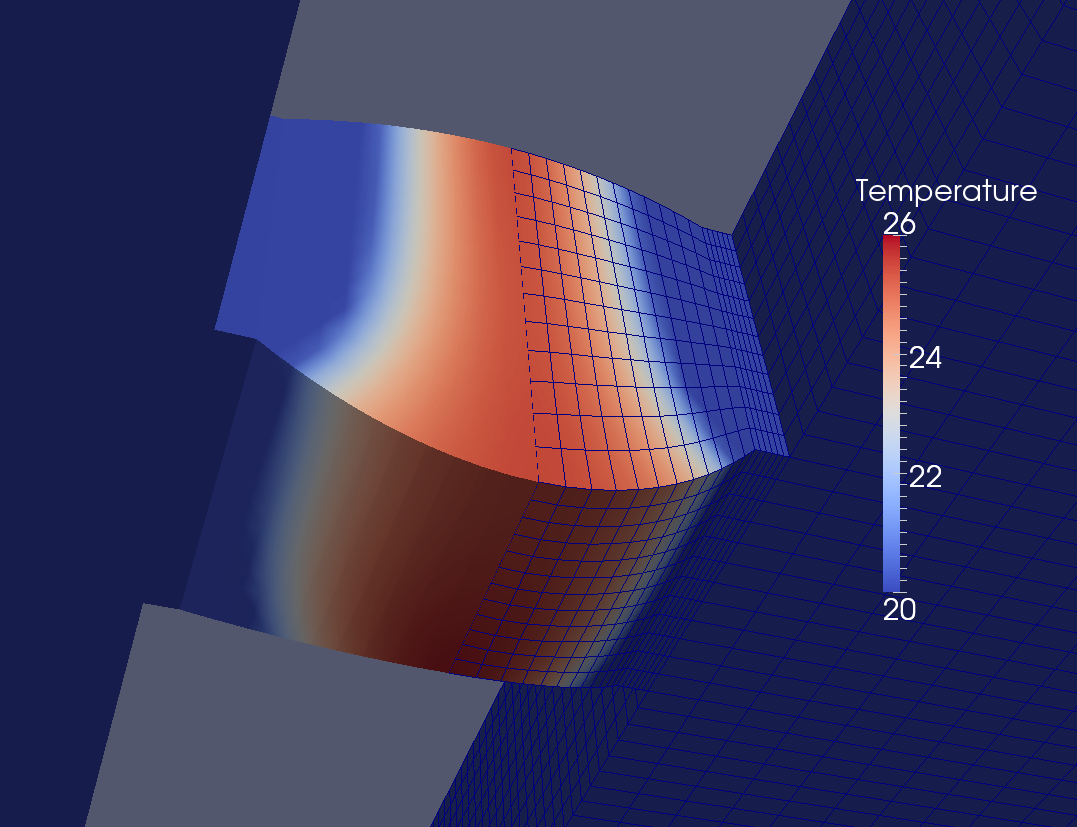}
\caption{Initial mesh configuration and steady-state temperature profiles
for the two-parameter optimization
problem. The deflections of the top and bottom surface of the slider are varied
independently, and the deflection (bulge) out of the plane is chosen to constrain
the volume to that of a rectangular brick.}
\label{fig:init2} \end{center} \end{figure}

\begin{figure}[h] \begin{center}
\includegraphics[width=4.2in]{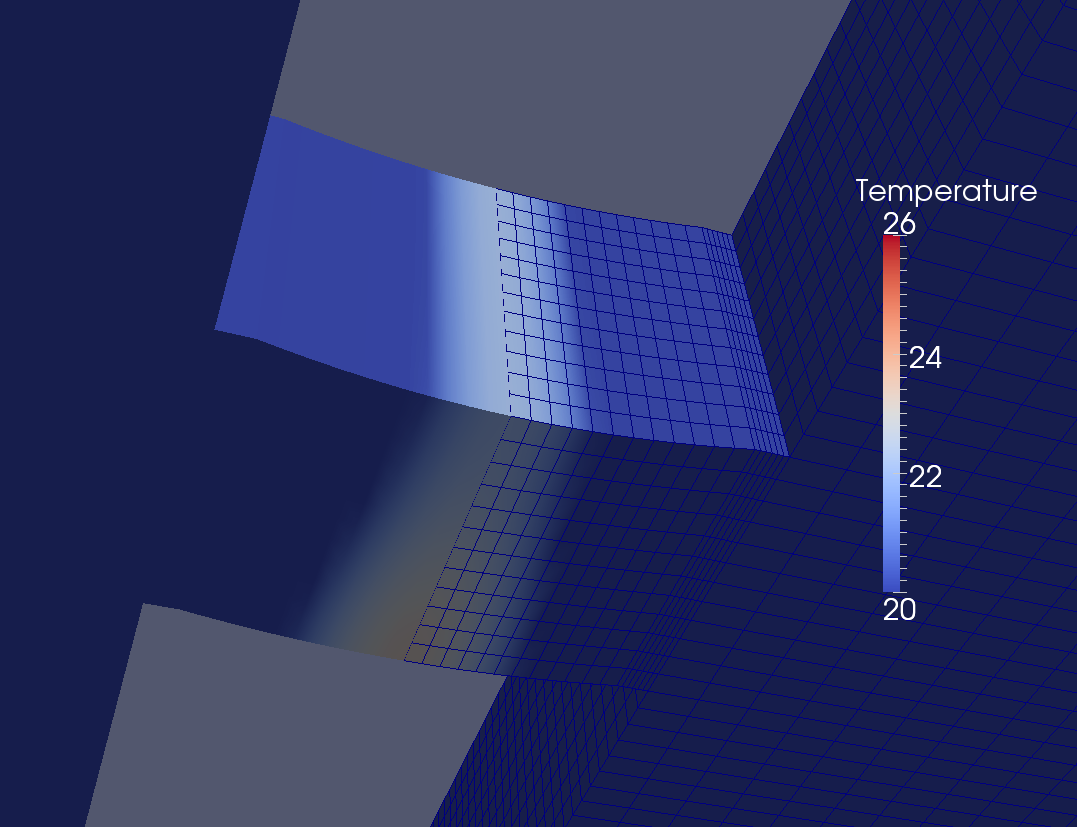}
\caption{Final configuration and color map for the two-parameter optimization
problem. The maximum temperature was significantly decreased.}
\label{fig:opt2} \end{center} \end{figure}

\subsection{UQ Results}

In addition to optimization runs, embedded uncertainty quantification was
performed on the same model. The \texttt{ScalarT} data type for the stochastic Galerkin
residual evaluation
hold polynomial coefficients for the expansion of all quantities with a
spectral basis. By nesting the stochastic Galerkin and Automatic Differentiation
data types, a Jacobian for the stochastic Galerkin expansions can also be
calculated. The Seed and Extract phases of this computation, as well as the
subsequent nonlinear solve of the stochastic FEM system required significant
development. However, with the template-based generic programming approach,
this work is completely orthogonal to the implementation of the PDEs. So,
there was no additional coding needed to perform embedded UQ for this
application over that needed for the
residuals for the governing equations ~(\ref{eq:gov1}--\ref{eq:gov2}).

As a demonstration, we chose the electrical conductivity in the
pad ${\sigma^p_0}$ as the
uncertain variable. The pad region is the thin rectangular region at the
edge of the slider with fixed shape. In this run, ${\sigma^p_0}$ was
chosen to be a uniform distribution within $15$ of the nominal value
of $35$,
\begin{eqnarray}
{\sigma^p_0}= [35.0 P_{0}(\xi) + 15.0 P_{1}(\xi)].
\end{eqnarray}
Here, the $P$ variables are Legendre polynomials. The computation
was run with degree-3 polynomial basis. A Newton iteration was performed
on the nonlinear system from the discretized 4D domain (3D FEM in space
+ 1 Stochastic Dimension with a spectral basis). The resulting probability
distribution on the maximum temperature unknown was computed to be
\begin{eqnarray}
T_{Max}= 25.87 P_{0}(\xi) + 0.61 P_{1}(\xi) - 0.17 P_2(\xi) + 0.04 P_3(\xi).
\end{eqnarray}
Figure~\ref{fig:mean} shows the mean temperature profile for this
distribution. Figure~\ref{fig:variation} show the variation of the
temperature with respect to this parameter. (Note the reduced range
of the color bar.) The results show that the variation of the electrical
conductivity in the pad region has a large effect on the temperature
in the middle of the slider (which has a large dependence on the total current),
and not in the pad region itself (which is strongly controlled
by the convective cooling from the beam due to the moving frame of reference).

\begin{figure}[h] \begin{center}
\includegraphics[width=4.2in]{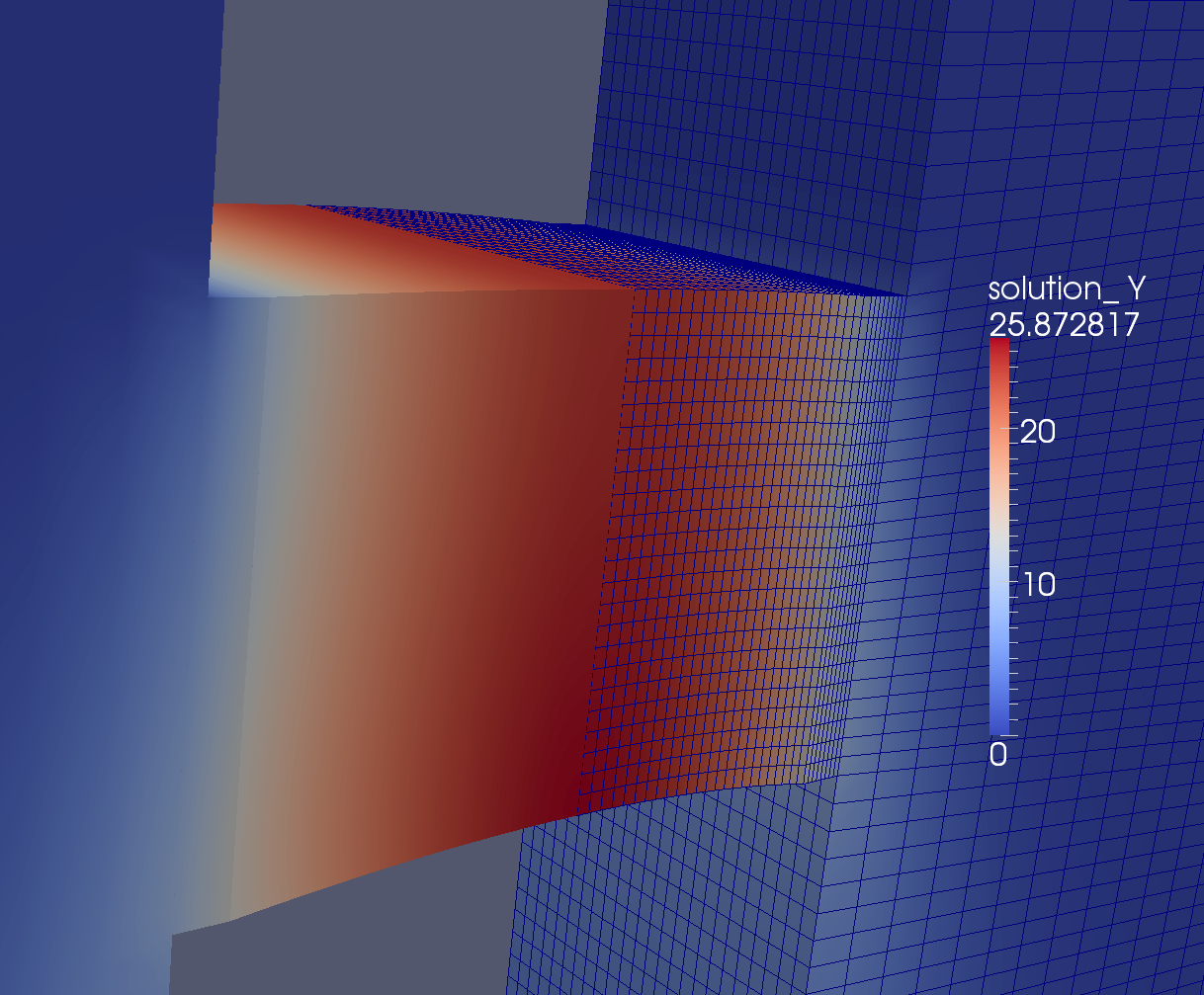}
\caption{The results of an embedded uncertainty quantification using Stokhos.
The electrical conductivity in the thin pad region is given as
a distribution. In this figure, the temperature profile for the mean solution
is shown.}
\label{fig:mean} \end{center} \end{figure}

\begin{figure}[h] \begin{center}
\includegraphics[width=4.2in]{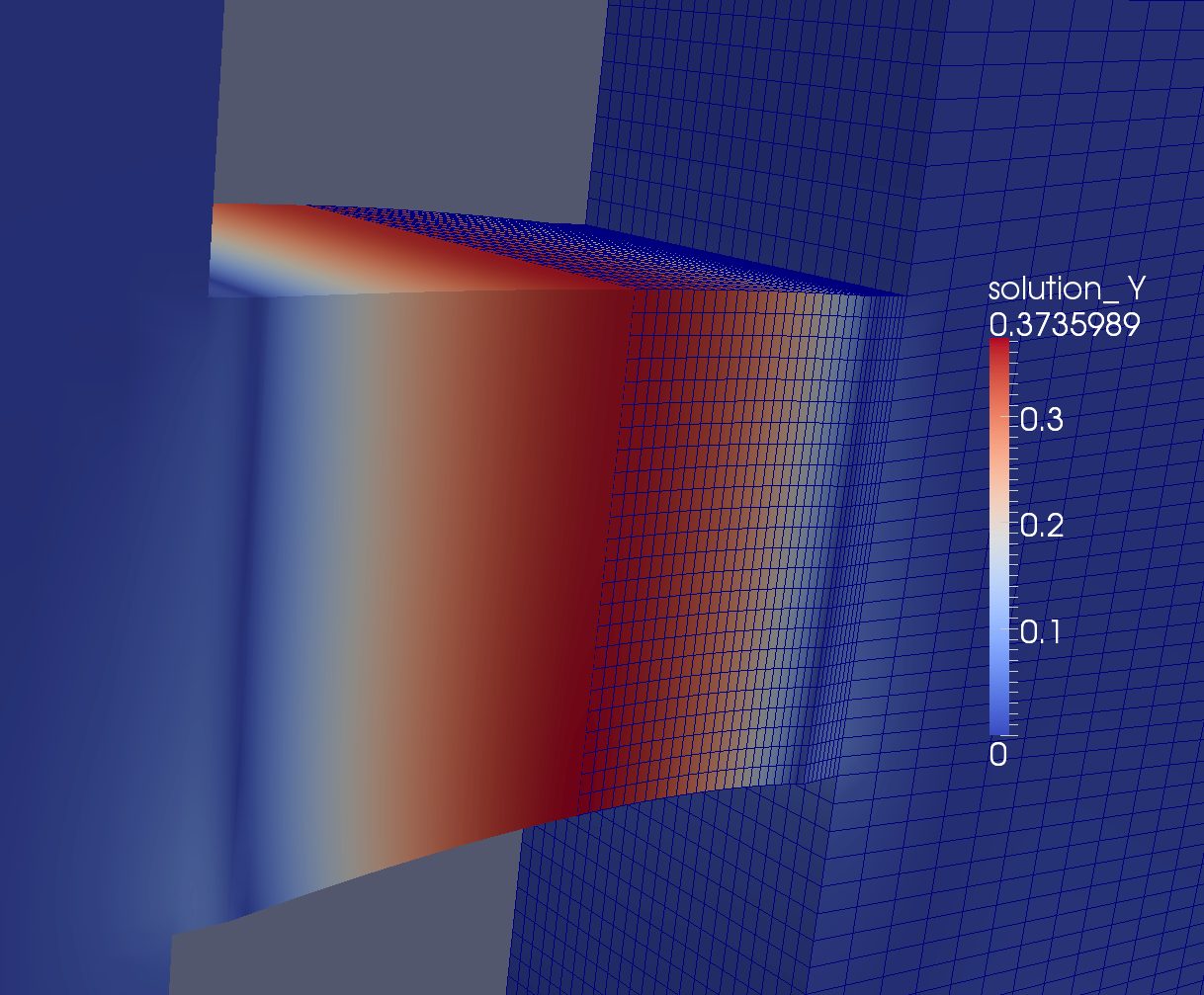}
\caption{The variation in the temperature field is shown, for the same
embedded uncertainty quantification calculation as the previous figure.}
\label{fig:variation} \end{center} \end{figure}

\section{Conclusions}

In this paper, we have related our experience in using the template-based
generic programming (TBGP) approach for PDEs in a finite element code.
We have used this approach
at a local element level, where the dependencies for Jacobian
evaluations are small and dense. By combining the Gather phase
of a finite element calculation with the Seed phase of the TBGP
approach, and the Scatter with Extract, the infrastructure
for TBGP is well contained. Once this infrastructure is in place,
transformational analysis capabilities such as optimization and
embedded UQ are immediately available for any new PDE modes that
are implemented.

We have also presented some of the implementation details in our
approach. This includes infrastructure for dealing with parameters,
for dealing with a templated code stack, and dealing with
non-templated code.
We demonstrated this approach on an example sliding electromagnetic
contact problem, which is a pair of coupled nonlinear steady-state equations.
We performed optimization algorithms with embedded gradients, and
also embedded Stochastic Finite Element calculations.

As this paper is to appear in a special issue along with other Trilinos
capabilities, we would like to mention explicitly which Trilinos packages (underlined) were used
in these calculations. This paper centered on the use of \underline{Phalanx}
assembly engine, \underline{Sacado} for automatic differentiation, and \underline{Stokhos} for
 embedded UQ. For linear algebra, we used the \underline{Epetra} data structures,
\underline{Ifpack} preconditioners, and \underline{Belos} iterative solver. The linear algebra was accessed through the \underline{Stratimikos} linear algebra strategy layer using the \underline{Thyra} abstraction layer.
The \underline{STK Mesh} package was
used for the parallel mesh database,
and the \underline{STK IO} packages, together with \underline{Ioss}, \underline{Exodus}, and \underline{SEACAS} were used for IO and partitioning of the mesh. The \underline{Intrepid} package was used for the finite element discretization,
operating on multidimensional arrays from the \underline{Shards} package.
The utility packages \underline{Teuchos} was used for parameter list specification and reference-counted memory management.

The \underline{Piro} package managed the solver and analysis algorithms, and
makes heavy use of the \underline{EpetraExt} Model Evaluator abstraction. Piro in turn calls
the \underline{NOX} nonlinear solver, the \underline{LOCA} library of continuation algorithms, the \underline{TriKota} interface
to the Dakota optimization algorithms, and \underline{Stokhos}
for presenting the stochastic Galerkin system as a single nonlinear
problem. These results also relied on several products outside
of Trilinos, including the Cubit mesh generator and associated mesh morphing software,
the Dakota framework, ParaView visualization package and netcdf mesh I/O library.

\section*{Acknowledgements}

This work was funded by the US Department of Energy through the NNSA Advanced Scientific Computing and Office of Science Advanced Scientific Computing Research programs.

\bibliographystyle{abbrvnat}
\bibliography{paperpde}

\end{document}